\RequirePackage[2020-02-02]{latexrelease}
\documentclass{aa}  

\usepackage{pgf}
\newcommand{\mathdefault}[1][]{}
\usepackage{graphicx}
\usepackage{txfonts}

\usepackage{aas_macros}

\usepackage[switch]{lineno}

\RequirePackage[utf8]{inputenc} 
\RequirePackage[T1]{fontenc}

\PassOptionsToPackage{dvipsnames}{xcolor} 
\usepackage{graphicx,pstricks}
\usepackage{color}
\graphicspath{{figures/}}

\usepackage{url}
\usepackage{xspace}

\usepackage{natbib}
\bibpunct{(}{)}{;}{a}{,}{,}

\usepackage[colorlinks=true,allcolors=blue]{hyperref}

\usepackage[nameinlink]{cleveref}

\usepackage{catoptions}

\makeatletter

\def\Autoref#1{%
  \begingroup
  \edef\reserved@a{\cpttrimspaces{#1}}%
  \ifcsndefTF{r@#1}{%
    \xaftercsname{\expandafter\testreftype\@fourthoffive}
      {r@\reserved@a}.\\{#1}%
  }{%
    \ref{#1}%
  }%
  \endgroup
}
\def\testreftype#1.#2\\#3{%
  \ifcsndefTF{#1autorefname}{%
    \def\reserved@a##1##2\@nil{%
      \uppercase{\def\ref@name{##1}}%
      \csn@edef{#1autorefname}{\ref@name##2}%
      \autoref{#3}%
    }%
    \reserved@a#1\@nil
  }{%
    \autoref{#3}%
  }%
}

\makeatother

\newcommand{\nuna}[2]{(#1)\,#2} 
\newcommand{\source}[1]{\textsuperscript{\textcolor{blue}{[citation needed]}}\xspace}

\newcommand{\numb}[1]{\textcolor{orange}{#1}}
\renewcommand{\numb}[1]{#1}

\newcommand{\add}[1]{#1}
\newcommand{\rev}[1]{{#1}}

\newcommand{\tess}{\texttt{TESS}\xspace}
\newcommand{\tsys}{\texttt{TSSYS-DR1}\xspace}

\newcommand{\ssocard}{\texttt{ssoCard}\xspace}

\newcommand{\ssodnet}{\texttt{SsODNet}\xspace}
\newcommand{\rocks}{\texttt{rocks}\xspace}

\newcommand{\palnbperiod}{9912}

\newcommand{\nbSSO}{9912}
\newcommand{\nbValid}{4839} 
\newcommand{\nbTenPercent}{4521} 

\newcommand{\nbUniquePer}{4366} 
\newcommand{\nbMultiPer}{492} 
\newcommand{\nbMultiOne}{277} 
\newcommand{\nbMultiN}{215} 
\newcommand{\fracValid}{48.8}
\newcommand{\fracUniquePer}{44.0} 
\newcommand{\fracMultiPer}{5.0} 
\newcommand{\fracMultiOne}{56.3} 
\newcommand{\fracMultiN}{43.7}  

\newcommand{\nbComparePal}{4829}
\newcommand{\nbCompareLit}{1483}
\newcommand{\nbCompareLitUnique}{670}
\newcommand{\fracSamePal}{78.5}

\newcommand{\fracAddPal}{3.6}
\newcommand{\fracSameLit}{91.0}

\newcommand{\fracAddLit}{0.9}

\newcommand{\nbCompareMcn}{6635}
\newcommand{\fracValidMcn}{40.0}
\newcommand{\nbCompareMcnVal}{3601}

\newcommand{\fracSameMcn}{63.1}

\newcommand{\fracAddMcn}{2.9}

\begin{document}

   \title{Rotation periods of asteroids from light curves of \tess
   data\thanks{The catalog is available at the CDS via anonymous ftp to
\url{cdsarc.u-strasbg.fr} (\url{130.79.128.5}) or via
\url{http://cdsarc.u-strasbg.fr/viz-bin/cat/J/A+A/XXX/xxx}}}


   \author{D. E. Vavilov
          \inst{1,2}
          \and
          B. Carry\inst{1}
          }

   \institute{Universit{\'e} C{\^o}te d'Azur, Observatoire de la C{\^o}te d'Azur, CNRS, Laboratoire Lagrange, France\\
             \email{benoit.carry@oca.eu}
         \and
         Institute of Applied Astronomy, Russian Academy of Sciences,
              Kutuzova emb. 10, St. Petersburg, Russia\\
              \email{vavilov@iaaras.ru}        
             }

   \date{Received September 15, 1996; accepted March 16, 1997}

 
  \abstract
   {Understanding the dynamical evolution of asteroids through the 
   secular Yarkovsky effect requires the determination of many physical
   properties, including the rotation period.}
   {We propose a method aimed at obtaining a robust determination of the rotation period
   of asteroids, while avoiding the pitfalls of aliases. We applied this approach to thousands of
   asteroid light curves measured by the NASA \tess mission.}
   {We developed a robust period-analysis algorithm based on a Fourier series.
   Our approach includes a comparison of the results from multiple orders and
   tests on the number of extremes to identify and reject potential aliases. 
   We also provide the uncertainty interval for the result as well as additional  periods that may be plausible.
   }
   {We report the rotation period for \numb{\nbTenPercent} asteroids within a
    precision of 10\%.
    A comparison with the literature (whenever available) reveals a very good agreement and validates the approach presented here.
    Our approach also highlights cases for which the determination
    of the period should be considered invalid.
    The dataset presented here confirms the apparent small number of asteroids
    with a rotation between 50 and 100 h and correlated with diameter.
    The amplitude of the light curves is found to increase toward
    smaller diameters, as asteroids become less and less spherical.
    Finally, there is a systematic difference between the broad
    C and S complex in the amplitude-period, revealing the statistically
    lower density of C-types compared to S-type asteroids.
   }
   {Our approach to the determination of asteroid rotation period
   is based on simple concepts, yet it is nonetheless robust. It can be applied to large corpora of
   time series photometry, such as those extracted from exoplanet
   transit surveys.}


   \maketitle
%
\section{Introduction}

The asteroids forming the Main Belt between Mars and Jupiter
are remnants of the bricks that accreted to form the planets.
Prints of the events that occurred in the early Solar System
are still present in the distribution of their orbit, size, and composition
\citep{2014Natur.505..629D, 2015-AsteroidsIV-Morbidelli, 2020MNRAS.492L..56C}.
The current population, however, differs from its pristine distribution.
Giga-years of collisions have fragmented bodies and created clumps of objects
called families \citep{1918AJ.....31..185H, 1990AJ....100.2030Z, 2014Icar..239...46M}.
All dynamical structures are furthermore secularly spreading through
the non-gravitational Yarkovsky effect 
\citep{2001Sci...294.1693B}.

The Yarkosky effect results from the delayed re-emission of the Solar incident
flux and depends on many physical and surface properties such
as  diameter,  albedo,  density,  obliquity,  and rotation period
\citep{2015-AsteroidsIV-Vokrouhlicky}.
Some properties, such as the diameter, are available for hundreds of
thousands of asteroids
thanks to mid-infrared (MIR) surveys such as IRAS, AKARI, or WISE
\citep{2002AJ....123.1056T, 2011PASJ...63.1117U, 2011ApJ...741...68M}.
Others, such as the density or thermal inertia, are much less constrained
and available for only a tiny fraction of asteroids
\citep[see][for a recent compilation]{ssodnet}.

Spin properties (rotation period and coordinates) typically require
numerous photometric measurements over a long period of time, covering
several apparitions \citep{2015-AsteroidsIV-Durech}.
These measurements can be dense-in-time time series, hereafter referred to as
"light curves" \citep{2001Icar..153...37K}, or they can be photometry sparse-in-time, collected
by surveys \citep{2004A&A...422L..39K}.
However, owing to the potential multiple period aliases from sparse data,
a light curve often helps to unambiguously determine  the rotation period
\citep{2015-AsteroidsIV-Durech}.

Surveys aimed at discovering and characterizing exoplanets via the transit method 
offer a tremendous amount of time series over wide fields 
in which asteroids can be searched for 
\citep{2016MNRAS.458.3394B, 2017-ACM-Grice}.
Recently, \citet{Pal_2020} extracted light curves
for \numb{\nbSSO} Solar system objects (SSOs) from 
the Transiting Exoplanet Survey Satellite (\tess)
first data release (DR1). The released catalog of asteroid
light curves and periods
is dubbed \tsys.
\citet{Pal_2020} determined the rotation period of \numb{\palnbperiod} asteroids by
fitting second order Fourier series.
However, in some cases, such as asteroids
(118) Peitho,
(511) Davida, and
(775) Lumiere among many others,
it is
not enough to describe accurately the
light curve as second-order in the Fourier Series \citep{2009Icar..200..531S},
leading to potentially erroneous period determinations.

In this work, we aim here to analyse the large sample of light curves
released by \citet{Pal_2020} with
a robust period-determination algorithm. We present examples
demonstrating that a Fourier series of the second order is
often not sufficient to describe asteroid light curves. 
The article is organized as follows.
In \Autoref{sec:fit}, we describe how we model the light curve of asteroids. In \Autoref{sec:select}, we explain how we selected the optimum solution among 
different potentially degenerated solutions.
We then present the results of the analysis of the \tess light curves
in \Autoref{sec:results}. We
validate these results in  \Autoref{sec:valid}
and we  discuss their implications
in \Autoref{sec:disc}.
Finally, we give our conclusions in Section 7. 

\section{Modeling asteroid light curves\label{sec:fit}}

The apparent visual magnitude of an asteroid depends on the
geometry of observation and can be described as \citep{1989aste.conf..524B}:

\begin{equation}
    V = H + 5 \log_{10} r \Delta - 2.5 \log_{10} \phi(\alpha) + g(t),
    \label{eq:Vis_mag}
\end{equation}

\noindent where $H$ is the absolute magnitude of the asteroid,
$r$ and $\Delta$ are the heliocentric distance and range to the observer (in au),
$g(t)$ is a periodic function related to the asteroid shape in rotation,
and the phase function $\phi(\alpha)$ describes how brightness evolves with the
phase angle $\alpha$
\citep[the Sun-asteroid-observer angle; see][for the different functions
in usage in the community]{1989aste.conf..524B,2010Icar..209..542M}.
The present study focuses on the determination of the period of the
$g(t)$ function.

\begin{figure}[t]
\centering
  \includegraphics[width=\hsize]{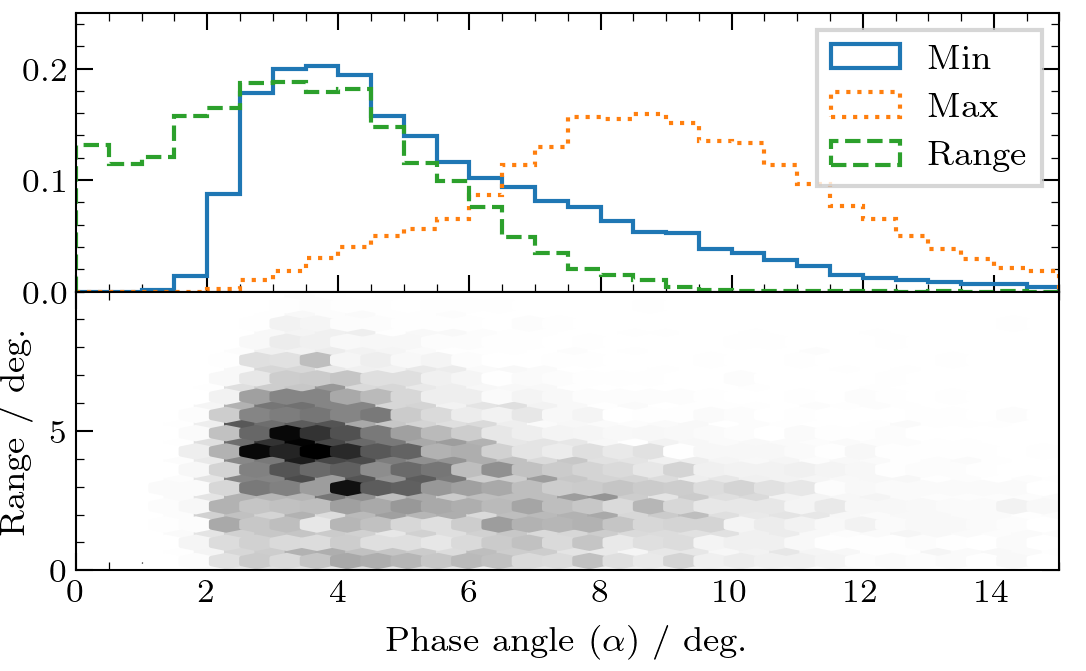}
  \caption{\textbf{Top.}  Distribution of the maximal, minimal, 
    and range of phase angles.
    \textbf{Bottom.} Range of phase as a function of the minimum phase angle. 
    \label{fig:phase_angles}
  }
\end{figure}

\subsection{Reduction to unit distances}

To find the rotation period of an asteroid,
we first take into account the change of visual magnitude 
occasioned by the changing distances due to the asteroid
and Earth orbital motions. For each observation, $i,$ we compute:

\begin{equation}
V'_i = V_i - 5 \log_{10} r_i \Delta_i .
\end{equation}

\noindent which depends only on the asteroid absolute 
magnitude, rotation, and phase function.
We also correct the timing of the $i$-th observation by:

\begin{equation}
t = t' - \frac{c}{\Delta_i}
,\end{equation}

\noindent where $c$ is the speed of light, $t'$ is the recorded observation epoch, and $t$ is the actual epoch for the observed visual magnitude.

Before moving into further analysis, we exclude the observations deemed to
be unreliable.
We exclude all observations with flags different from 0
\citep[see Table~1 in][for a full description of the flags]{Pal_2020}.
Observations with flag values 128 and 16384 
(manual exclusion for anomaly or because outlier)
were only excluded
from the first iteration of fitting and may be included
in the next iterations according to a $3\,\sigma$ rule.


\begin{figure}[t]
\centering
  \input{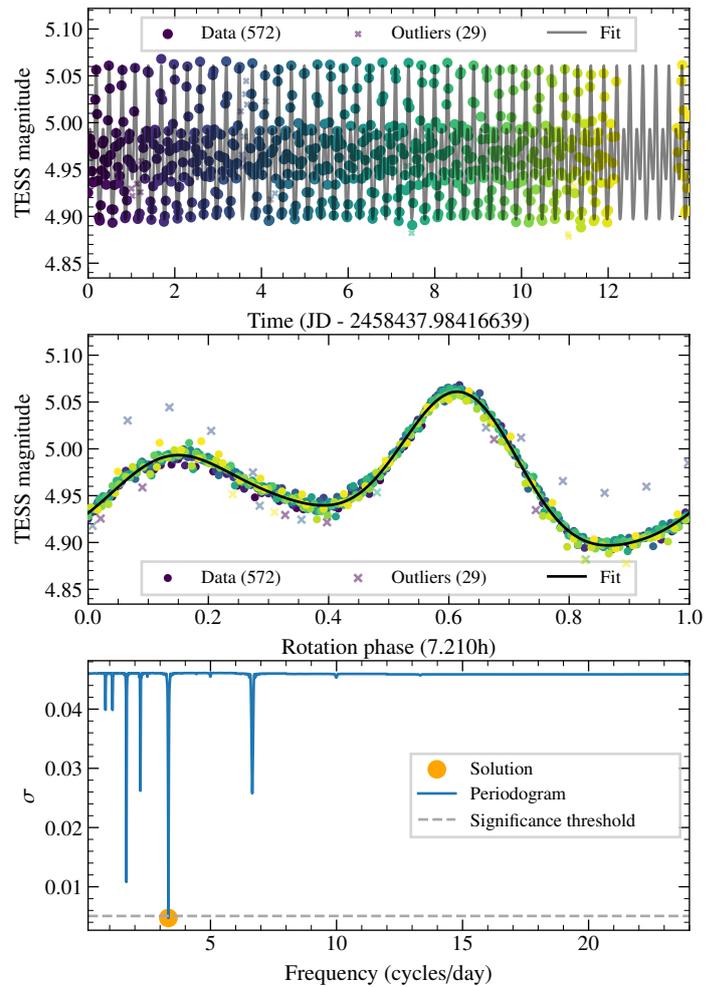}
  \caption{Example of light curve fit for asteroid (3) Juno.
  \textbf{Top.} \tsys data as function of time. Observations are plotted as dots, color-coded by
  epoch. Crosses represent data rejected by the fitting procedure (see text).
  The line represents the best fit.
  \textbf{Middle.} Same as above, folded over a rotation period
  ($7.210_{ -0.001 }^{ 0.001 }$).
  \textbf{Bottom.} Periodogram showing the residuals of the fit as a function of  
  rotation frequency (number of cycles per day). \rev{The horizontal
  dashed line correspond to the significance threshold
  (sigma-level, see text).}
  }
     \label{fig:example}
\end{figure}

\subsection{Fourier fitting and phase function}

The reduced $V'$ magnitude fluctuates because of the phase function $\phi(\alpha)$
and the periodic function $g(t)$, with a period equals to the asteroid's synodic rotation period.
The function $g(t)$ is decomposed into a Fourier series:

\begin{equation}
    g(t) = \sum_{j=1}^k A_j \cos (2\pi f j t) + B_j \sin (2\pi f j t),
\end{equation}

\noindent where $f$ is the frequency of rotation (cycles per day),
$t$ is time (in days),
$k$ is the maximal number of Fourier series, and
$\{A_j,B_j\}$ the amplitude coefficients at each frequency.
In the present analysis, we consider values of $k$ up to $k=10$.

The phase angles of asteroids observed by \tess typically span a small range
or a few degrees only, around 3\degr~of phase, with minimum values
of typically 3\degr~and maximum values around 8\degr~(\Autoref{fig:phase_angles}).
The lack of coverage of the opposition surge
implies that the system of equations is ill-conditioned for 
finding the parameters of the usual phase functions
$H,G$ \citep{1989aste.conf..524B}
or
$H,G_1,G_2$ \citep{2010Icar..209..542M}.
We refer to \citet{2021Icar..35414094M} for a discussion of 
phase coverage.
We hence follow here the approach by \citet{Pal_2020}
of describing the phase function as a second-degree polynomial:

\begin{displaymath}
\phi(\alpha) = c_0 + c_1 \alpha + c_2 \alpha^2 .
\end{displaymath}

\begin{figure*}[ht]
\centering
\begin{tabular}{cc}
  \input{figures/fig_outcomes.pgf}
\end{tabular}
\caption{
  Illustrations of the advantages of the method (symbols are the same as
  Fig.~\ref{fig:example}).
  \textbf{Left:} Clear period determination, but it would be (erroneously) twice
  smaller if only the second order of the Fourier series was used.
  \textbf{Center:} Two solutions with acceptable sigma values but corresponding to the 
  same period -- if the number of local maxima is taken into account.
  \textbf{Right:} All solutions below the significance level, thus, the uncertainty
  interval is 2--237\,h, i.e., the period is not determined.
  }
\label{fig:outcomes}%
\end{figure*}

This approach precludes the determination of the absolute magnitude, $H,$ of an asteroid,
but that is not our goal.
On the other hand, the fitting is accurate and well conditioned.
We limit here the possible range of values for $c_1$ and $c_2$ to ascertain the physical results
(i.e., increasing reduced magnitude with phase angle).  
We used the $H,G$ phase function \citep{1989aste.conf..524B}
as a reference and assumed that the parameter $G$ cannot take values outside the given range $(-0.25, 0.95)$. 
\add{The H-G function cannot be properly approximated by a second degree polynomial. That is why we find an approximation on the arc of $1\degr$.}
For each angle $\alpha$ from $\{ 1\degr, 2\degr, 3\degr,..., 120\degr \}$
we construct two phase curves with ($G_{max} = 0.95$ and $G_{min} = -0.25$).
We then fit parameters $c_0$, $c_1$, and $c_2$ on the arc $(\alpha-0.5\degr, \alpha+0.5\degr)$.
The maximal and minimal values for $c_1$ and $c_2$ found this way for a particular phase angle
are then used as boundaries for the fit of \tess data.
We should note here that the free term $c_0$ encompasses $H$, which is removed from the fitting procedure. 

For a given frequency, $f$,
we found the parameters $c_0$, $c_1$, $c_2$, $\left\{A_j, B_j\right\}_{j = 1,k}$
via a least-squares minimisation.
We took the weights of observations into account according to chapter VIII in \citet{Linnik1961}.
We also excluded observations differing from the fit by more than 3\,$\sigma$ \add{iteratively until the procedure converged}. 

We put ``soft'' limitations on the parameters $c_1$ and $c_2$ in the least-squares fitting
procedure, \add{so that these values cannot be substantially outside the interval $[c_j^{max}, c_j^{min}]$. } \add{We added} the following two equations to the system of conditional equations \citep[p. 73]{Gubanov1997}:

\begin{equation} 
c_j = (c_j^{max}(\alpha) + c_j^{min}(\alpha)) / 2, \ j = 1,2, 
\end{equation}

\noindent where $\alpha$ is a minimal phase angle in the observation data of the asteroid.
These equations are added with weights of 
$1 \left/ (c_j^{max}(\alpha) - c_j^{min}(\alpha)) / 2 \right.$.
The aim of it is to guide the fitting procedure toward mean values of $c_1$ and $c_2$
of $(c_j^{max}(\alpha) + c_j^{min}(\alpha)) / 2$,
with standard deviations of $(c_j^{max}(\alpha) - c_j^{min}(\alpha)) / 2$.
We present in \Autoref{fig:example} an example of the fitting procedure, on asteroid
\nuna{3}{Juno}.

\section{Selection of the best model\label{sec:select}}

To find the rotational period, we sampled different frequencies, $f$, 
and for each of them we fit via least-squares routine the $2k + 3$ parameters 
$c_0$, $c_1$, $c_2$, $\left\{A_j, B_j\right\}_{j = 1,k}$.
We did not simply select the solution with the lowest residuals; rather, we imposed the following criteria to be fulfilled.

\subsection{Defining the frequency range} 

First, at least two full periods of the asteroid rotation must have been observed.
Shorter coverage could indeed result in spurious period determination.
We thus set a minimal frequency:
\begin{equation}
  f_{min} = \frac{2}{t_{max}-t_{min}}    
,\end{equation}
\noindent where $t_{max}$ and $t_{min}$ are the epochs of the last and first observation, respectively. 
Owing to the \tess observing strategy, each sector is observed for approximately a month, namely: $f_{min} > 0.05\ days^{-1}$

Furthermore, \tess full-frame images are taken every 30\, min approximately \citep{Pal_2020}.
The maximal possible frequency is thus analogous to a Nyquist-limit, adapted to the case of irregular sampling
\citep{1999A&AS..135....1E}: where $p$ is the largest value, such that each $t_i$ can be written as $t_i = t_0 + n_i p$, for integers, $n_i$; then the Nyquist frequency is $f_{Ny} = 1/(2p)$. Of course, the equality $t_i = t_0 + n_i p$ cannot be fulfilled precisely, so we allowed for a 2\% accuracy for this equation.
For most asteroids in \tsys, the Nyquist frequency is approximately 24~days$^{-1}$, 
implying a minimal rotational period of about 1 hour (however, in some cases, $f_{Ny}$ reached about 48~days$^{-1}$). We set
the right end of the frequency interval to 24~days$^{-1}$, thus: a minimal rotation period of 1 hour.

We note that the maximal frequency decreases with the number of Fourier series (i.e., greater $k$).
The $k$ series contains a term in  $\cos (2\pi f k t)$ which period is $1/kf$, and hence the maximal frequency is $f_{Ny} / k$. 
We consider 40,001 possible frequencies on the interval between $f_{min}$ and $min(f_{Ny}, 24)$ for $k=1$.
For $k>1$, we keep the frequency steps and the interval becomes $[f_{min}, min(f_{Ny}/k, 24)]$.

\subsection{Choosing the optimum fit}

For each number of Fourier series, $k$,
we first found the best fit according to our criteria (\Autoref{eq:sigma}).
We then compared fits obtained with different $k$ to select the final solution. 

The main criterion is the minimal value of standard deviation of one observation $\sigma$,
which is a classical unbiased estimate of a observational dispersion \citep{Linnik1961}:

\begin{equation}
    \sigma^2 = \frac{N_{obs}}{\sum_j p_j}   \frac{\sum_i p_i (O_i - C_i)^2 }{(N_{incl}-n_{par})},
    \label{eq:sigma}
\end{equation}

\noindent where $O_i$ and $C_i$ are $i$-th observation and computed magnitude,
$p_i$ is the weight of $i$-th observation (taken as 1/$\sigma_i^2$),
$N_{obs}$ is the number of observations,
$N_{incl}$ is the number of observations included in the fit, and
$n_{par}$ is the number of fitted parameters ($2k + 3$).

We note that the number of observations included in the fit can slightly differ for each frequency.
The factor $N_{obs}$ in $N_{obs} \left/ \sum_j p_j \right.$ is required
to bring the value $\sigma$ to the observational error with mean weight.
This does not change the result but helps us understand the accuracy of the fit.
We chose the model with the lowest value of $\sigma$ as solution for this number of Fourier series.

If the shape of an asteroid is well-described by a tri-axial ellipsoid, the light-curve 
is expected to display two local maxima and two local minima. However, the number 
of local maxima from fitting can be up to $k$. For instance, 
eight local minima can appear with eight Fourier series.
This point is critical as many objects present two 
solutions for the frequency, only differing by a factor of two. 
There are cases where the
two solutions are in reality associated with the same period
(e.g., \Autoref{fig:outcomes}), particularly if one of the solutions
refers to a light curve with one local maxima only.



We repeated the above procedure for each number of Fourier series, $k$, from 1 to 10.
For each $k,$ we have a single candidate for the final model and final period. 
To choose among these, we used the F-test and computed it for each pair of models.
\add{The F-test tells us whether the difference between two dispersions 
\rev{is} significant or not. The F-test gives us the probability (p-value) for the first model being better than the second one. }
In total, we obtained $10 \times 9/2 = 45$ results for the F-tests.
Following the Occam's razor rule, we chose the model with smallest \add{possible} number of Fourier series.
We chose the smallest $k$ for which no other model is significantly better, namely, \add{ such a $k$ value that all the p-values of F-tests of $k$ and any $j$ are less than $95\%$}

For this chosen model, we computed its sigma-level (F-test, p-value $90\%$). \add{With the F-test, we computed $p_{value} = F_{test}(\sigma_1, \sigma_2, n_1,n_2)$. However, we can also solve the reverse problem: \rev{for a given $\sigma_1$, the aim is to find the $\sigma_2$ of the second model,} so that the p-value is $90\%$. In this case, we used the same number of used observations $(n_1=n_2)$. This is how we define the sigma-level.}
All the models with $\sigma$ values lower than this threshold are also possible, with their
associated rotation period.
As a result, we provide a range of valid rotation period and, whenever applicable, 
all the other potential periods that are not in the computed interval
(i.e., generally a multiple of 2 of the period).
In the case of a high number of possible periods, we provide the whole interval for all, which means that the result is imprecise.

Also, we checked whether using a different number of Fourier series can lead to any plausible solutions (the $\sigma$ value of which is lower than for the chosen one). In general, this is the case when for a higher order of Fourier series, the frequency can  only be low (so the period is high) and the period is doubled or even tripled. 
These periods might be realistic (i.e., the $\sigma$ is still lower than for chosen solution), but still unlikely.


\section{Results\label{sec:results}}

\begin{figure}[t]
\centering
\input{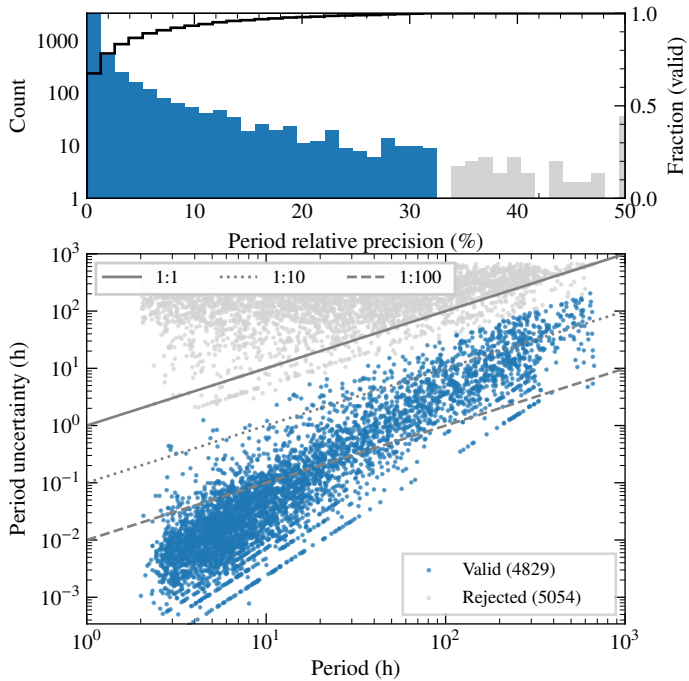}
  \caption{Comparison of the uncertainty on the periods with 
    the periods.
    \label{fig:per_err}
    }
\end{figure}

  From the sample of \numb{\nbSSO} asteroids with \tess light curves, 
  we found a period for all, with
  \numb{\nbValid} (\numb{\fracValid\%}) for which the period
  has been determined 
  \add{with a $33\%$ relative error}
  (i.e., the uncertainty on
  the period is at most a third of the period).
  Hereafter, we consider as ``valid'' the periods determined
  with this level of precision and ``reject'' the others.
  While there is a long tail
  of less precise determinations (\Autoref{fig:per_err}), the 
  determinations are more precise than a percent for
  \numb{3042} (\numb{30.7\%}) asteroids
  and 10\% for
  \numb{\nbTenPercent} (\numb{45.6\%}) asteroids.
    
\begin{figure}[t]
\centering
\input{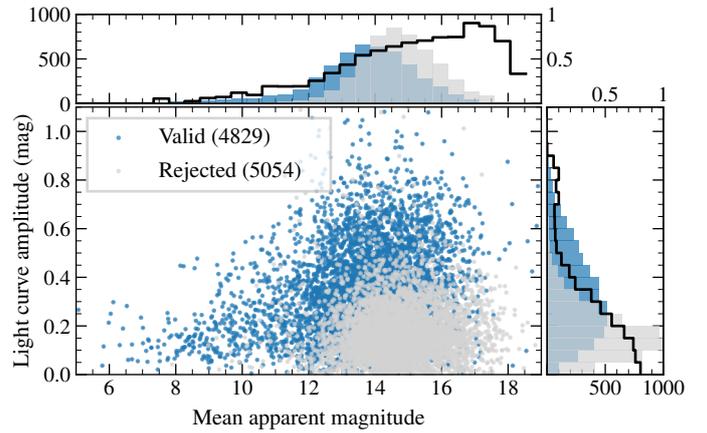}
  \caption{Distribution of the validated and rejected samples
    against light curve amplitude and apparent magnitude.
    The black lines represent the fraction of rejected asteroids
    in the entire sample.
    \label{fig:mag}
  }
\end{figure}

  We present in \Autoref{fig:mag} the distribution of the 
  validated and rejected solutions as a function of the 
  apparent magnitude of the asteroids during \tess observations
  and the amplitude of their light curves.
  There is a clear trend with rejected solutions being mostly for
  the faintest targets with low amplitude light curves.
  Hence, inaccurate
  solutions are due to a level of noise that is too high compared
  to the signal, as expected.

\begin{figure}[t]
\centering
\input{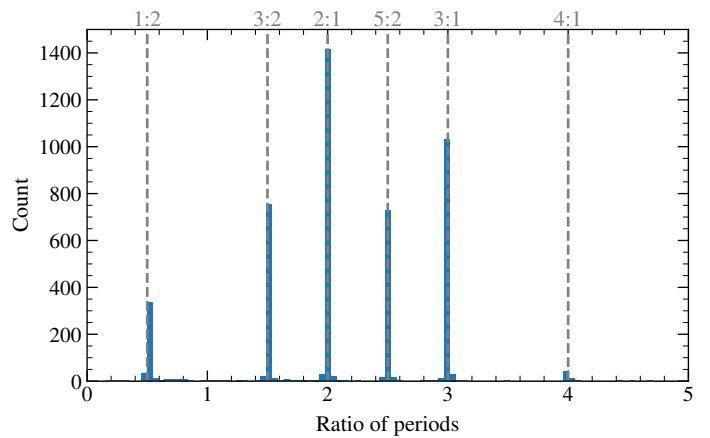}
  \caption{Ratio of periods between ambiguous solutions.
    The vertical dashed lines indicate the ratio of integers.
    \label{fig:alias}
  }
\end{figure}

  Finally, \numb{\nbUniquePer} (\fracUniquePer\%) asteroids have a unique,
  non-ambiguous, rotation period.
  The remaining \numb{\nbMultiPer} (\fracMultiPer\%) asteroids have a single
  other possible period in 
  \numb{\nbMultiOne} (\fracMultiOne\%) cases, and several ambiguous periods in
  \numb{\nbMultiN} (\fracMultiN\%) cases.
  The distribution of these degenerated solution is not random but corresponds
  to period aliases, as visible in \Autoref{fig:alias}.
  All the data are available as electronic format on the CDS.

\section{Validation\label{sec:valid}}

\begin{figure*}[t]
  \input{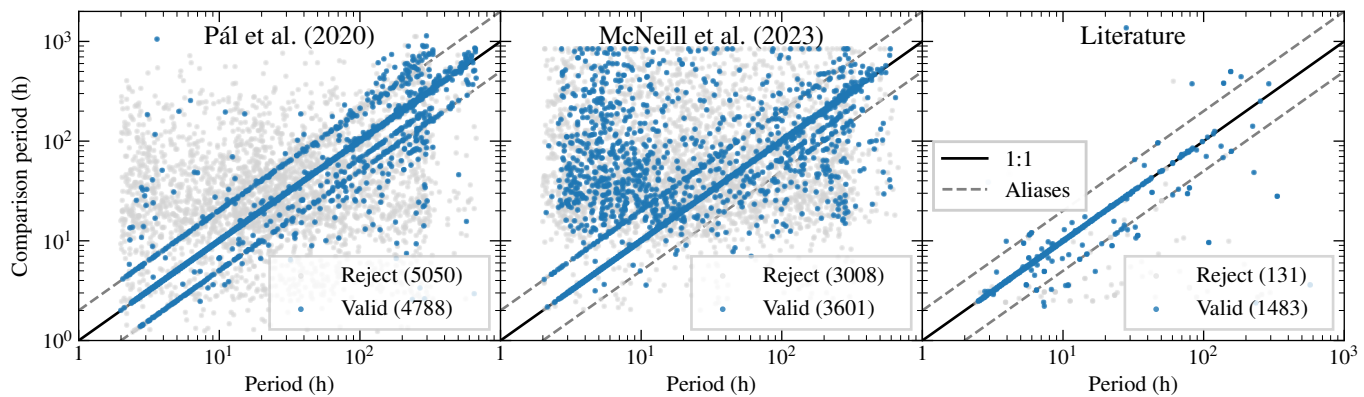}
  \caption{Comparison of rotation periods found here with 
         those derived by \citet{Pal_2020} (\numb{\nbSSO}, \textsl{left})
         \add{by \citet{McNeill2023} (\numb{\nbCompareMcn}, \textsl{center})}
         and gathered from the literature (\numb{\nbCompareLit} periods
         for \numb{\nbCompareLitUnique} asteroids, \textsl{right}).
         The black line represents full agreement and the dashed gray
         lines the aliases for double and half periods.
          \label{fig:period_compa}
          }
\end{figure*}
%

  We assessed the quality of the periods determined here with \add{three} comparisons:
  the original \tsys results of \citet{Pal_2020},
  \add{the rotation periods
  on an independent extraction of 
  \tess photometry by \citet{McNeill2023}},
    and other values from the literature \add{(providing periods from altogether different
    data sets)}.
  \add{As visible in \Autoref{fig:period_compa}, there is a significant spread among
  periods when comparing the present determinations with
  those from \citet{Pal_2020}. It, however, mainly concerns rejected periods.}
  Among the validated
  solutions (\numb{\nbComparePal}, i.e., \numb{\fracValid}\% of the sample),
  the comparison is much better and
  \numb{\fracSamePal}\% of the solutions agree within 1\%.
  Another \numb{\fracAddPal}\% correspond to period aliases (half or double period).
  \add{The comparison with \citet{McNeill2023} presents a larger spread, even
  in the validated 
  solutions (\numb{\nbCompareMcnVal}, i.e., \numb{\fracValidMcn}\% of the common sample).
  Overall, \numb{\fracSameMcn}\% of the solutions agree within one percent
  and an additional \numb{\fracAddMcn}\% correspond to period aliases.
  While based on similar original \tess data, the two studies extracted the
  photometry and determined the period using different methods, explaining
  the differences observed here.
  In their study, \citet{McNeill2023} actually reported numerous cases of
  disagreement with \citet{Pal_2020}, while about 80\% of solutions agreed.
  For instance, while mainly periods reported by \citet{Pal_2020} correspond
  to either half or twice the period reported here (\Cref{fig:period_compa}),
  those reported by \citet{McNeill2023} almost never correspond to half the
  period reported here.
  }

  We then made a comparison with periods reported by other authors on different data sets
  (excluding \tess).
  We thus compiled the rotation period for each asteroid
  in our sample, using the \ssocard of 
  \ssodnet\footnote{\url{https://ssp.imcce.fr/webservices/ssodnet/}} through its 
  \rocks\footnote{\url{https://rocks.readthedocs.io/en/latest/}} interface
  \citep{ssodnet}. 
  \add{We focus on period determinations with a quality flag of 3, 
  that is, those that are deemed definitive, following the criterion by LCDB
  \citep{2009Icar..202..134W}.}
  As visible on \Autoref{fig:period_compa}, there is globally a 
  better agreement with \numb{\nbCompareLit} period
  determinations (for \numb{\nbCompareLitUnique} unique asteroids).
  We find \numb{\fracSameLit}\% agreeing within one percent,
  with additional \numb{\fracAddLit}\%
  corresponding to aliases.
  This highlights the robustness of the method used here to select the
  period among the different possible solution (\Autoref{sec:select}).

  There are three main advantages of the technique presented in the present study.
  First, we considered several number of Fourier series, while also taking into account the number of local maxima 
  (there should be at least two local maxima, otherwise we would be doubling the period). \add{We assume that the light curve is a result of the shape features.}
  In \Autoref{fig:outcomes} one can see the result for asteroid (511) Davida.
  The period is clearly determined and in agreement with
  many previous studies from ground-based light curve observations
  \citep[e.g.,][]{1995P&SS...43..649D, 2003Icar..164..346T,
  2019A&A...631A..67C, 2021A&A...654A..56V};
  however, at least the third order of a Fourier series is required. 
  The result for the second order is twice smaller and coincides 
  with \citet{Pal_2020}. Our explanation for this is that the 
  light curve is quite $\pi$-periodic, on the one hand, 
  and it does not resemble a cosine, on the other hand.
  Therefore the second-order Fourier series with twice smaller period can fit the contour of the light curve much better, but once we take third-order Fourier series, the difference between the two peaks of the light curve becomes important.
  
  Second, we checked the number of local extrema.
  We illustrate this second advantage with 
  asteroid (699) Heia in \Autoref{fig:outcomes}.
  The lower part of the figure shows that there is two possible 
  solutions with radically different frequencies. 
  However, the second frequency of 14.13 cycles/day provides
  only one local maxima and hence its period should be multiplied by 2,
  yielding the same fit to the data.
  
  Finally, we did not just provide the chosen solution, 
  but all the possible solutions.
  Sometimes there can be two or three possible solutions
  (periods for which $\sigma$ is not significantly higher). 
  These periods are also possible and we report their values. 
  For some asteroids, the data are not accurate enough 
  to obtain reliable results and the range of possible
  rotational periods ends up including the entire possible interval: 
  from 2\,h to 360\,h.
  This interval helps us understand the quality and reliability of the result.
  In the case of asteroid 2014 AX12 (\Autoref{fig:outcomes}),
  the uncertainty in the photometry is so high that almost any period
  can be fitted. The uncertainty interval for this asteroid is [2\,h, 237\,h].
  This basically means that the result should not be considered,
  even though there is a formal solution associated with the minimal value of $\sigma$.


\section{Discussion\label{sec:disc}}

\begin{figure}[t]
  \input{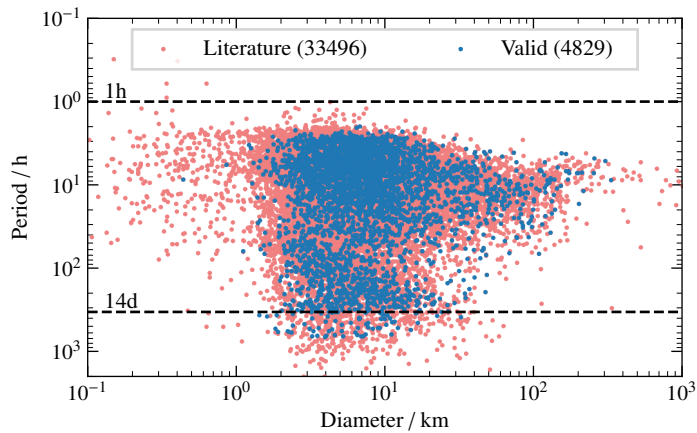}
  \caption{Distribution of the rotation period
    as function of the asteroid diameter, compared with the literature.
    The two horizontal dashed lines represent the minimum and
    maximum periods that can be determined here (Sect.~\ref{sec:select}).}
     \label{fig:per_diam}
\end{figure}

We present in \Autoref{fig:per_diam} how periods derived here
are distributed against the  diameter measurements. We also present data compiled from the
literature \citep[retrieved from \ssodnet,][]{ssodnet}.
The limits imposed by \tess observations are clearly visible.
First, the cadence of exposures (30\,min) and length of 
observations (27\,days) limit the range of periods that can be determined
(\Autoref{sec:select}), between 1\,h and 27\,days.
Second, there is a clear drop in the amount of solutions below
a diameter of 2--3\,km. This diameter roughly corresponds to an
apparent magnitude of 18 in the asteroid belt, which is indeed the 
peak in the magnitude distribution in \tsys data set
\citep{Pal_2020}. This distribution, combined with the trend
for faintest asteroids to
have the highest rejection rate (\Autoref{fig:mag}), explains the
drop of solutions below 2--3\,km in diameter.

The lower limit of 1\,h is not reached by any of our solutions.
As also visible in the data from the literature, there is a clear boundary
at 2.2\,h which is refereed to as the ``spin barrier'' and corresponds
to the critical rotation period at which self gravity and centrifugal
acceleration are balanced \citep{2000Icar..148...12P}.
The situation is different for the upper limit, with longer
rotation periods reported in the literature from long-term campaigns
of observation or archival data
\citep[e.g.,][]{2015AJ....150...75W, 2021MNRAS.506.3872E, 2021A&A...654A..87M}.
We note the presence of a valley around 50--100\,h
separating the bulk of fast rotators from the slower ones, with
an apparent dependence on diameter.
This low-density region was already visible in the rotation periods
determined from K2 by \citet{2021ApJS..254....7K} and from
Gaia by \citet{2023A&A...675A..24D}.

\begin{figure}[t]
  \input{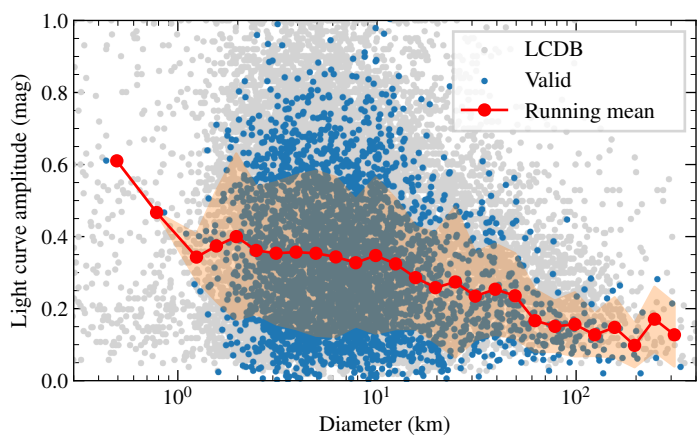}
  \caption{Distribution of the light-curve amplitude
    as a function of asteroid diameter. 
    The asteroids in LCDB \citep{2009Icar..202..134W} are plotted for 
    reference, as well as a running mean\rev{ (red line) and
    standard deviation (shaded area)}.
    \label{fig:amp_diam}
    }
\end{figure}

We present in \Autoref{fig:amp_diam} the distribution of
the amplitude of the light curves as a function of the asteroid diameter.
We also present the asteroids reported in the
LCDB \citep{2009Icar..202..134W} for comparison.
Most asteroids have a diameter between 3 and 20\,km.
There is a clear trend of higher light-curve amplitudes
toward smaller diameters,
revealed by the running average.
Such a trend was already visible in the results of the 
Palomar Transient Factory on asteroids 
\citep{2015AJ....150...75W}.
The amplitude of the light curve is directly related to the change 
of the projected area of the shape on the plane of the sky. Smaller
amplitudes are thus indicative of rounder shapes.
The trend here is a 
clear signature of asteroids being less spherical at smaller diameters.
Such a trend has been reported from the results of 3D shape modeling of the
42 of the largest (diameter above 100\,km) main belt asteroids,
with an increase of asphericity toward smaller diameters
\citep{2021A&A...654A..56V}.
The regular increase in amplitude below the 100\,km reveals that
the trend continues down to diameters as small as 2\,km (size
at which the present data set is limited).

\begin{figure}[t]
\centering
  \input{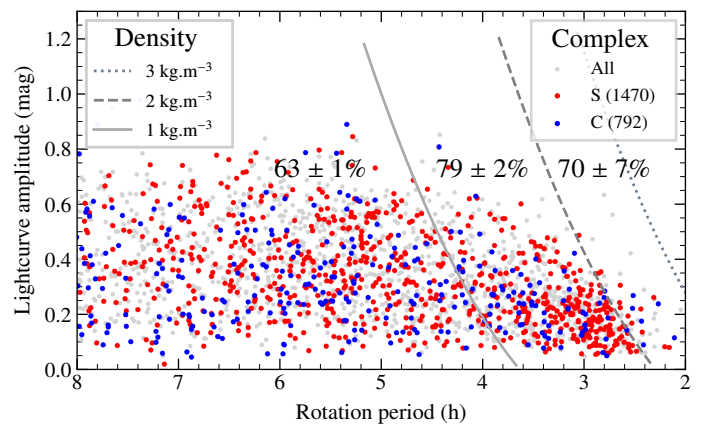}
  \caption{Distribution of amplitude against rotation period.
  The gray lines represent 
  the fastest spin for bodies held together solely
  by self-gravity from \citet{2015AJ....150...75W}.
  The fraction of S-types among all asteroids is indicated
  between those curves.
  }
  \label{fig:amp_period}
\end{figure}

We compare the light-curve amplitude to the rotation period in 
\Cref{fig:amp_period}, focusing on the shortest rotation periods.
The period distribution is not random, with a limit at the shortest  
periods dependent on the amplitude of the light curve.
As a guideline, we present  the theoretically largest
possible amplitude for bodies held together by self-gravity only
\citep[taken from,][]{2015AJ....150...75W}
for three reference densities: \add{1000, 2000, and 3000} kg$\cdot$m$^{-3}$.
There is a clear difference between asteroids in a broad ``S'' complex
(including the S, Q, A, V, E classes) and those in a
broad ``C'' complex \add{\citep[C, Ch, B, D, P, Z classes, see][for
a rationale on this grouping]{2022A&A...665A..26M}}, as expected
from the higher density of the first group.
The fraction of S-like asteroids is higher below the spin limit of
\add{1000}\,kg$\cdot$m$^{-3}$, revealing the lower density of C-like asteroids
(below or slightly above \add{1000}\,kg$\cdot$m$^{-3}$) compared to
S-like asteroids, closer to \add{2000}\,kg$\cdot$m$^{-3}$.

\section{Conclusion}

We present a new approach of determining the rotational period of asteroids
from optical light curves. The key aspects of the approach are: 

\begin{itemize}
\item Fit of Fourier series to the data by the least-squares method.

\item Selection of the model with the lowest weighted root-mean square residuals.

\item Selection of the lowest possible order of Fourier series, following the principle of "the less the better.''

\item Check with the F-test if higher orders Fourier series are necessary. 

\item Test the number of local maxima of the function. Multiply the period by two if only one local maxima is present (we assume that the shape mostly produces the light curve)

\item Report the determined rotation period with its uncertainties. This includes the confidence interval of the period and the possible alternative rotational periods (ambiguous solutions with similar residuals, generally alises of the solution). If there are too many alternative periods, an interval encompassing all these periods is reported.
\end{itemize}

This approach was used to compute the rotational periods of asteroids observed 
by Transiting Exoplanet Survey Satellite (\tess) reported by
\citet{Pal_2020}. This dataset has 
observations of \numb{\nbSSO} asteroids. We determined the
period of \numb{\nbTenPercent} asteroids with an accuracy better than 10\%. 

Comparison of our results with \citep{Pal_2020} shows \numb{\fracSamePal}\% of a full agreement.
For some of the asteroids, we identified the advantage of our technique,
in particular, when fitting higher order of Fourier series and checking for
the number of local maxima. 
The method we propose here is robust and can be applied to any dataset of dense light 
curves of asteroids. Thus, we plan to use it on available asteroid datasets in a future work.

\begin{acknowledgements}
   B. Carry acknowledges support by the French ANR, project T-ERC
   SolidRock (ANR-20-ERC8-0003).
   This paper includes data collected by the \tess
   mission. Funding for the \tess mission is provided by the
   NASA Explorer Program.
   We did an extensive use of the VO tools
   TOPCAT \citep{2005ASPC..347...29T}
   and \ssodnet \citep{ssodnet}.
   Thanks to all the developers and maintainers.
\end{acknowledgements}

  \bibliographystyle{aa} 
  \bibliography{bibliography,ssodnet} 

\end{document}